\documentclass[referee]{aa} 
\usepackage{graphicx}
\usepackage{txfonts}
\begin{document}
   \title{Observational searches for g-mode oscillations in the quiet solar atmosphere from TRACE 1600 $\AA$ 
Continuum Observations}

\author{R. Kariyappa
          \inst{1}
          \and
          L. Dam\'e\inst{2}}


   \institute{ Indian Institute of Astrophysics, Bangalore 560034, India\\
              \email{rkari@iiap.res.in}
         \and
             Service d'A\'eronomie du CNRS, BP \# 3, 91371 Verrieres-le-Buisson Cedex, France\\
             \email{luc.dame@aerov.jussieu.fr}}
\offprints{R. Kariyappa}
 
  \date{Received ........................ / Accepted ....................... }

 
  \abstract
{}  
   {Our aim is to search for atmospheric g-mode oscillations in UV network, UV  bright points and Uv background regions.}
   {We have analysed a 6-hours of time sequence of ultraviolet (uv) images
obtained on May 24, 2003 in 1600 $\AA$ continuum under high spatial and temporal
resolution from Transition Region and Coronal Explorer (TRACE).
We have selected an isolated 15 uv bright points, 15 uv network elements and
15 uv background regions in a quiet region from the images for the detailed analysis.}
{We derived the cumulative intensity values of these features.
The light curves of all the features have been derived for the total
duration of observations and done the power spectrum analysis using the time
series data.  We found that the uv bright points, the uv network and uv background
regions will exhibit a longer period of intensity oscillations namely, 5.5 hours, 4.6 hours
and 3.4 hours respectively, in addition to the more familiar small scale intensity
fluctuations.  We suggest that the longer periods of oscillation may be related to
solar atmospheric g-modes.}
   {}


\titlerunning {Observational searches for atmospheric g-mode oscillations}
\authorrunning {R. Kariyappa and L. Dam\'e}
   \maketitle
%

\section{Introduction}
The gravity waves play an important role in studying the coupling
of lower and upper solar atmospheric regions and are therefore of tremendous
inter-disciplinary interest.  The gravity waves in the Sun can be divided into
two types, namely, (i) the
internal gravity waves, may be confined to the solar interior, and (ii) the atmospheric
gravity waves, which are related to the
photosphere and chromosphere, and may be further beyond.  In general, the observation of gravity
mode oscillations of the Sun would provide
a wealth of information about the energy-generating region, which is poorly
probed by the p-mode oscillations.  In addition, the internal g-modes of the Sun are the
most powerful tool for the investigation of the solar core, and a way to solve,
for instance, the neutrino problem.  It has been suggested that the turbulent
convection below the photosphere will generate the high order, non-radial g-mode
oscillations (internal gravity waves)
(Meyer and Schmidt, 1967; Stix, 1970).   There are theoretical studies earlier on
solar-atmospheric
gravity waves by various groups (Whitaker, 1963, Lighthill, 1967, Stein, 1967 and
Schmieder, 1977).  In Frazier's (1968) {\it k - w} diagrams, the traces of
internal gravity waves may be present.  Deubner (1974) observed the generation
of internal gravity waves by individual granules.  Cram (1978) has investigated
the evidence of low, but significant,
power at frequencies relevant to internal gravity waves.  He had also concluded from the studies
of the phase lag between successive layers, that there was an
upward energy flux.  In addition, Brown and Harrison (1980) have observed the
indications of the possible existence of trapped gravity waves by analyzing the brightness
fluctuations of the visible continuum.  These internal gravity waves, by-products of the
granulation, are expected to be fairly common and may not be negligible in the energy balance
of the lower chromosphere.  Pall\'e (1991) had discussed in great detail on various methods
to search for solar gravity modes.

        In recent years, an increasing amount of attention has been given to the
possible effects of internal gravity waves in the interiors of the Sun and stars.  Such waves
are likely to be excited when convective down-flows in Sun's outer envelope penetrate into the
underlying stably stratified, radiative layers. The internal gravity waves were not observed
with that much evidence earlier and they can be
attributed to several reasons. As a result of strong radiative damping, the gravity
waves cannot propagate in the photosphere (Souffrin, 1966), and thus may not be observed in lines
formed in this region.  As pointed out by Deubner (1981), the gravity waves were
expected to be extremely difficult to observe because there are local, small-scale features
requiring very high spatial resolution observations.  High temporal \& spatial resolution data
will reveal that these gravity waves are small-scale phenomena.  Many authors have claimed, in
the past 22 years, to detect g-modes in the Sun, but, so far, there is no observational
evidence.  Using wavenumber and frequency-resolved (k,f)
phase-difference spectra and horizontal propagation diagram, Straus and Bonaccini (1997) have
presented observationally, the strongest evidence of gravity wave presence in the middle
photosphere.   There are some observational investigations to show that there is a signature of
atmospheric gravity waves at the chromospheric level
using the time sequence of filtergrams and spectra obtained in CaII H \& K and Mg b2 lines
(Dam\'e et al. 1984, Kneer and von Uexkull, 1993, Kariyappa, et al. 2006).  Recently, Rutten and
Krijger (2003) have analysed the ultraviolet (1700 $\AA$) and white-light image sequences of
internetwork regions from TRACE and shown that there
is a signature of atmospheric gravity waves.

        In the present paper, we made an attempt in search of atmospheric g-modes in the
lower choromosphere
using the long time sequence of intensity oscillations in a quiet region at the sites of the
uv bright points, uv
network elements and uv background regions observed under a high spatial, spectral and
temporal resolution in 1600 $\AA$
from TRACE Space Mission.  We will be presenting the first results of these analysis.


\section{Observations and Data Analysis}


We obtained a coordinated and simultaneous observations during May 18-24, 2003
with TRACE, SOHO/MDI and SOHO/CDS experiments.  A high spatial and temporal resolution of images
have been obtained almost at the center of the solar disk covering both active and quiet regions.
The solar rotation correction has been taken care during the observations.  The TRACE observations
are obtained in three wavelength regions: 1550 $\AA$, 1600 $\AA$ and 1700 $\AA$.   In this paper,
we have used the
observations obtained with TRACE on May 24, 2003 in 1600 $\AA$ UV continuum and it is a 6-hour
long time sequence of uv images.  These images have been analysed in IDL using SolarSoftWare (SSW).
For the preliminary study, we have chosen 15 uv bright points (UVBP), 15 uv network elements
(UVNW), and 15 uv background regions (UVBG) in a quiet region.
 We have used the square/rectangular boxes covering the selected features
for the study.  Then we have summed up all the pixel intensity values covered by the box and
extracted the cumulative intensity of a chosen feature for the entire 6-hours duration of
observations.  The light curves of all the UVBPs, UVNWs, and UVBGs have been derived and
plotted them as a function of time.  We have done a power spectrum analysis on the time series
data to determine the period of intensity oscillations associated with these features.


\section{Results and Discussion}

There was an indication of the existence of longer-period of oscillations in
chromospheric bright points and network elements from CaII H-line observations.
Since it was only a 35-minute duration of time sequence of observations, it was
difficult to investigate on the longer period of
oscillations (Kariyappa, et al. 2006).  In order to confirm on the existence of longer period of
oscillations, in this paper, we have analyzed
a long time sequence of uv images (6 hours of observations) obtained on May 24, 2003
with TRACE in 1600 $\AA$ UV continuum.  We identified and chosen 15 uv bright points (UVBPs),
15 uv network elements (UVNWs), and 15 uv background regions (UVBGs) from the time sequence of
uv images.  We derived the cumulative intensity values of the UVBPs, UVNWs, and UVBGs using SSW
in IDL.  To calculate the intensity we have put the rectangular or square boxes covering the
selected features. We derived the
intensity time series of all the ultraviolet bright points (UVBPs), uv network (UVNWs) and
uv background regions (UVBGs).  As an example we have shown the time series of the
two UVBPs (UVBP1 and UVBP2) from our selection in the upper panel of Fig.1.  The time series
of UVBPs show a small fluctuations in their intensity values.  In addition there is an
indication of longer period.  To determine the period of intensity oscillations, we have done
the power spectrum analysis using their time series data.
The power spectra for UVBP1 and UVBP2 are shown in the lower panel of Fig.1.  It is clearly
seen from the power spectra the existence of significant \& prominent peak around 5.5 hours in
both the cases.  Similarly, we have shown the time series and power spectra for two uv network
elements (UVNW1 and UVNW2) respectively in the upper and lower panels of Fig.2.  As we could
see from the power spectrum plots that the uv network elements exhibit around 4.6 hours of
period of intensity oscillations.  In the upper and lower panels of Fig.3, we have presented
the time series and power spectra for two background regions (UVBG1 and UVBG2).  The background
regions will be associated with around 3.4 hours of period of intensity oscillations.  We have
performed the cross spectrum analysis on the uv bright points and uv network elements of May 22,
2003 to compare with May 24, 2003 observations.  We found that both the
data sets show a coherent in phase \& there is a single dominate period associated
with uv bright points (around 5.5 hours) and uv network elements (4.6 hours).  It has high
coherence between May 22 and 24 uv bright point and uv network modulation.  This suggests
strongly for high-order
atmospheric gravity waves and they can be excited by turbulent stresses in the convection zone.

\begin{figure}
   \centering
  \includegraphics[width=8.0cm,height=7.5cm]{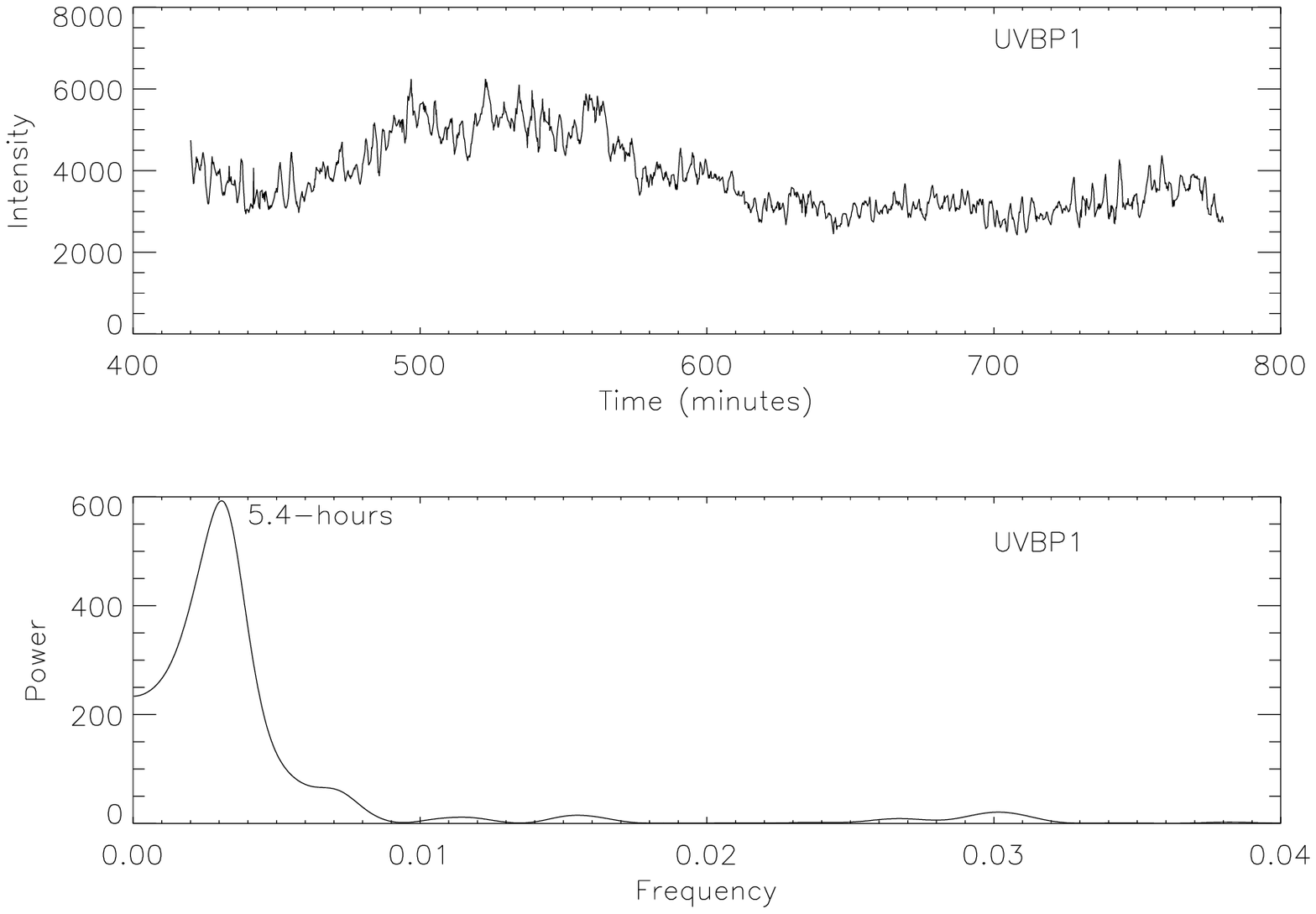}
  \includegraphics[width=8.0cm,height=7.5cm]{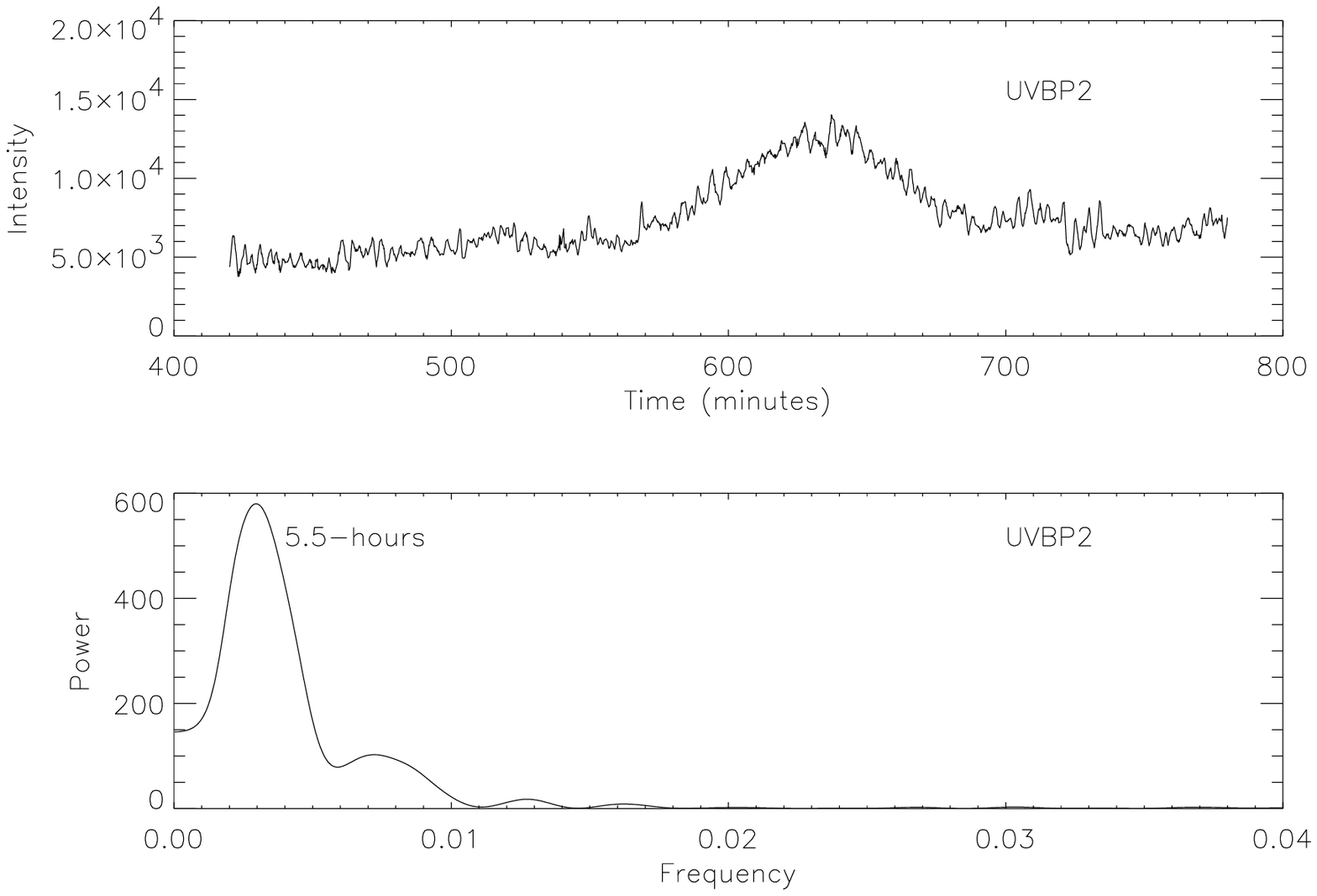}
\caption{{\it Left upper box:} An example of the light curve of an isolated UV bright point (UVBP1)
observed on May 24, 2003
(6 hours) with TRACE in 1600 $\AA$ UV continuum. {\it Left lower box:} The power spectra taken for
the light curve of the UV bright point (UVBP1).
{\it Right upper box:}  The light curve of an another isolated UV bright point (UVBP2).
{\it Right lower box:} The power spectra taken for the light curve of the UV bright point (UVBP2).}
\label{FigVibStab}
    \end{figure}

\begin{figure}
   \centering
  \includegraphics[width=8.0cm,height=7.5cm]{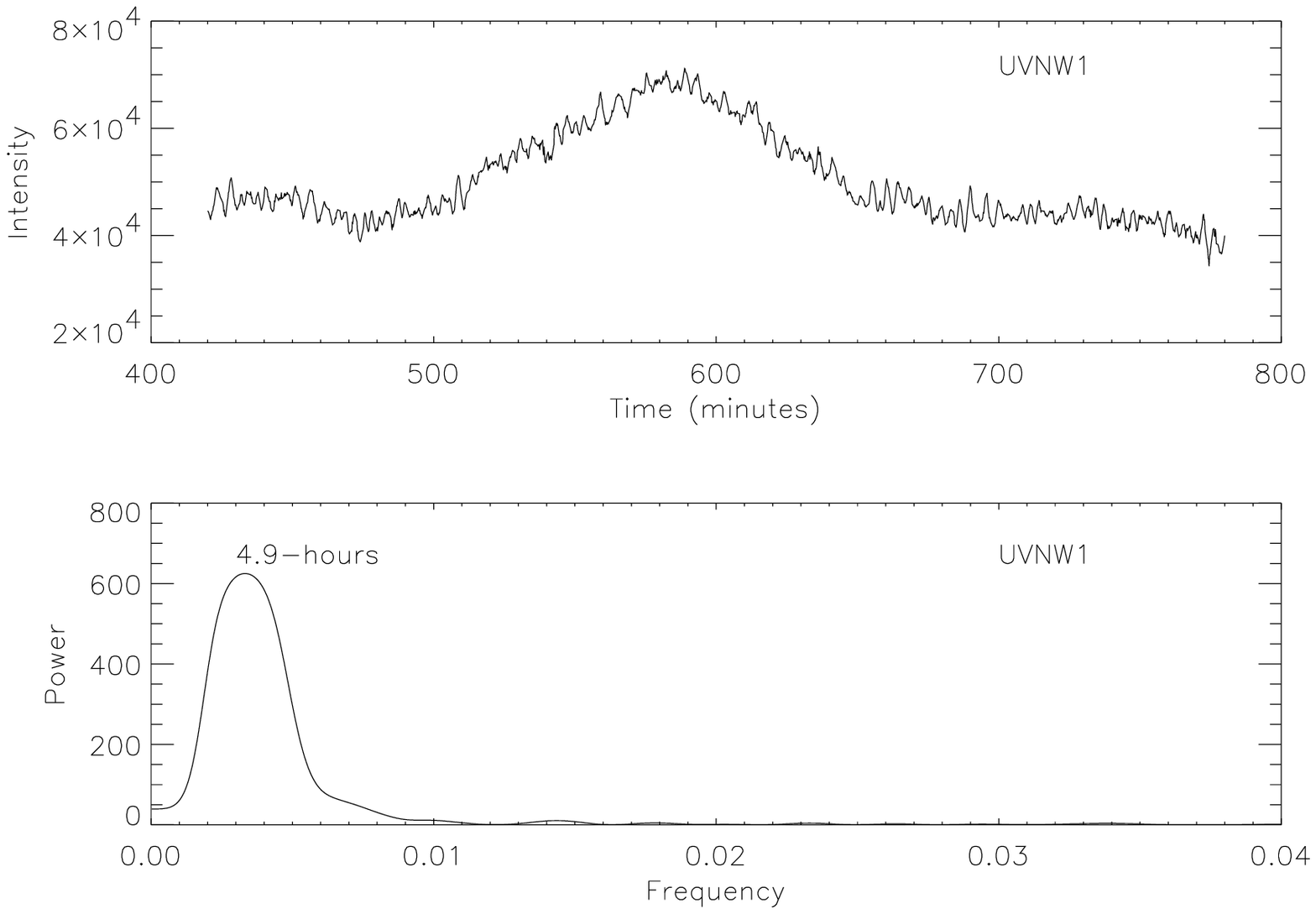}
  \includegraphics[width=8.0cm,height=7.5cm]{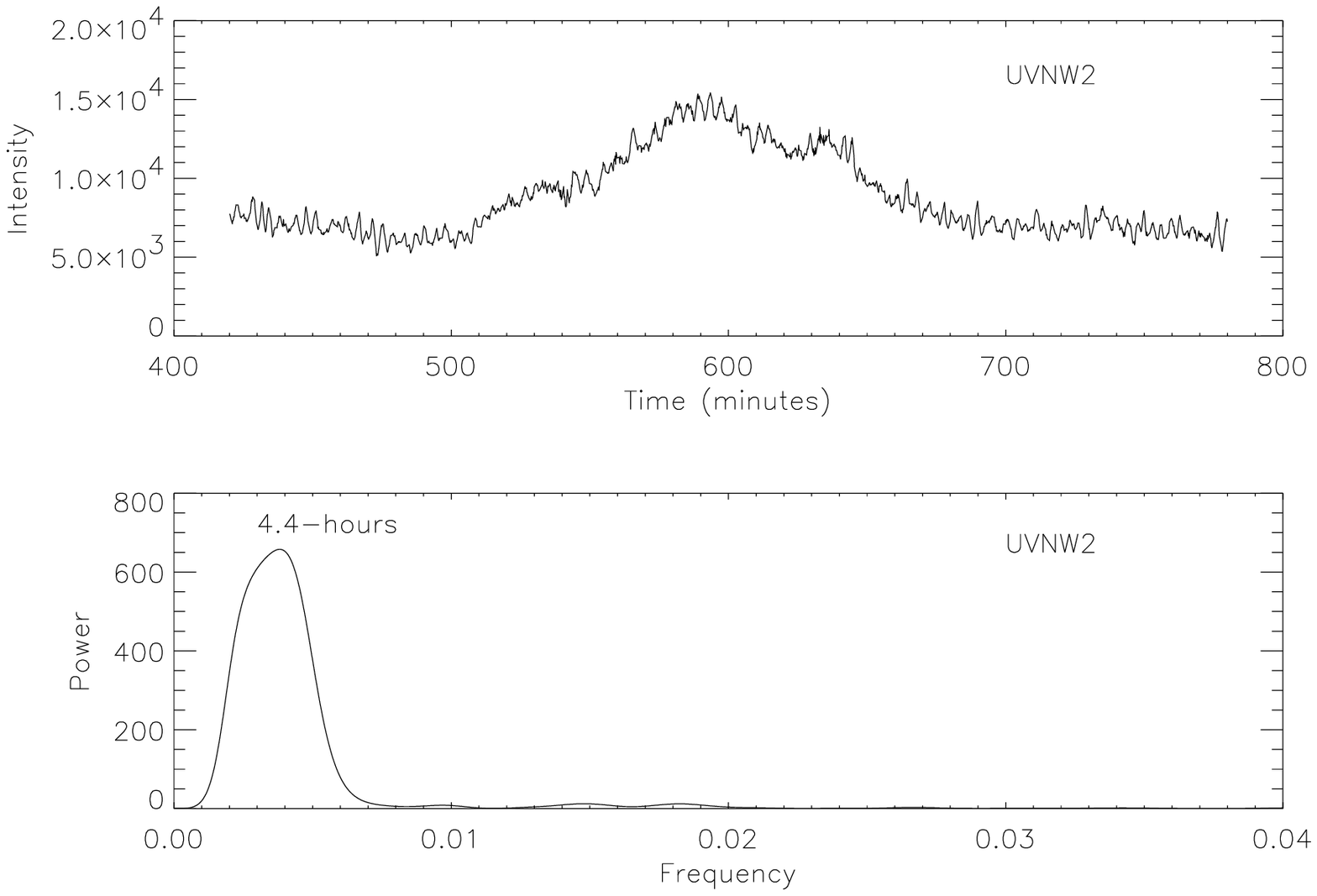}
\caption{{\it Left upper box:} An example of the light curve of a uv network element (UVNW1)
observed on May 24, 2003 (6 hours) with TRACE in 1600 $\AA$ UV continuum. {\it Left lower box:}
The power spectra taken for the light curve of the uv network (UVNW1).
{\it Right upper box:}  The light curve of an another uv network element (UVNW2).
{\it Right lower box:} The power spectra taken for the light curve of the uv network (UVNW2)}
\label{FigVibStab}
    \end{figure}

\begin{figure}
   \centering
  \includegraphics[width=8.0cm,height=7.5cm]{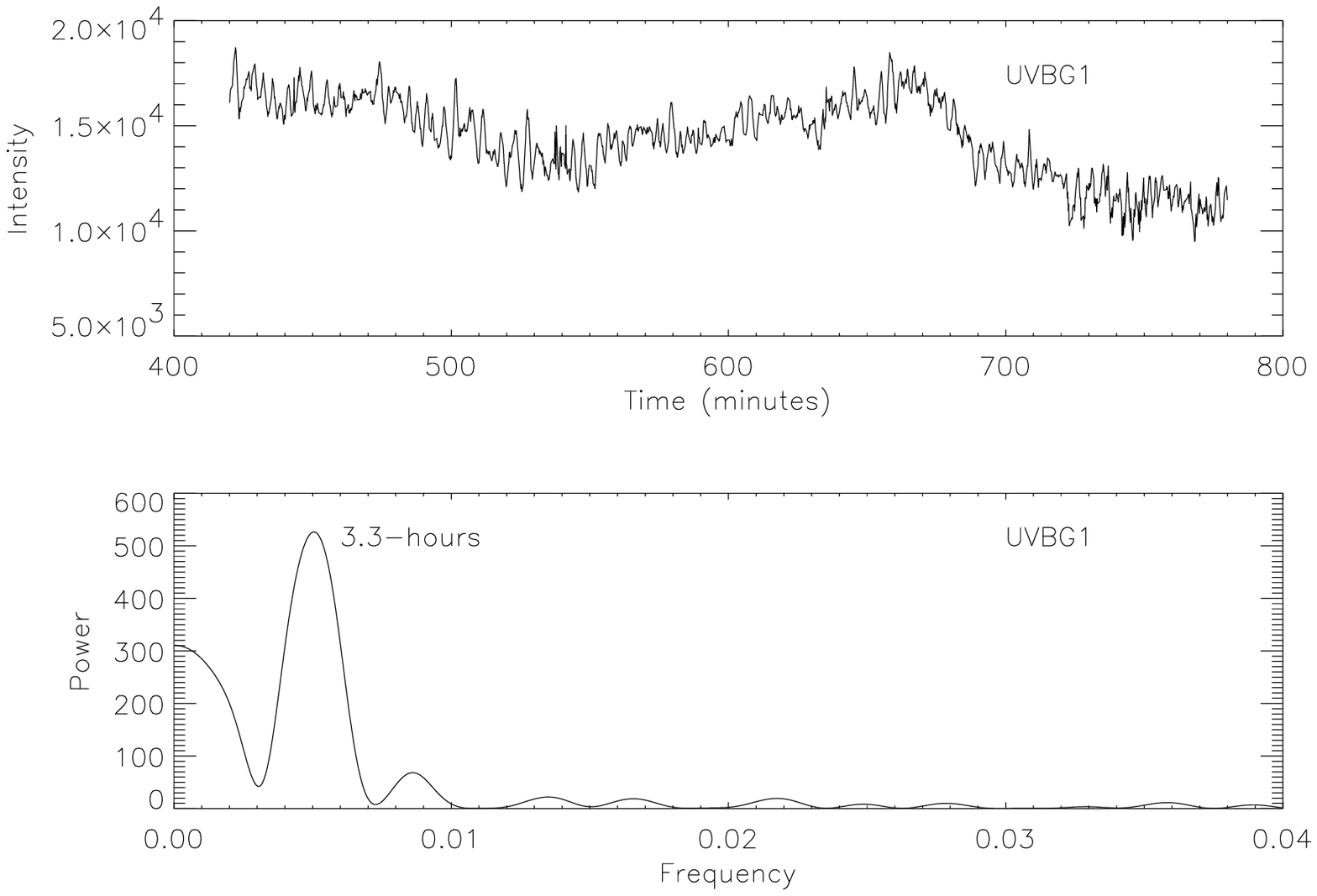}
  \includegraphics[width=8.0cm,height=7.5cm]{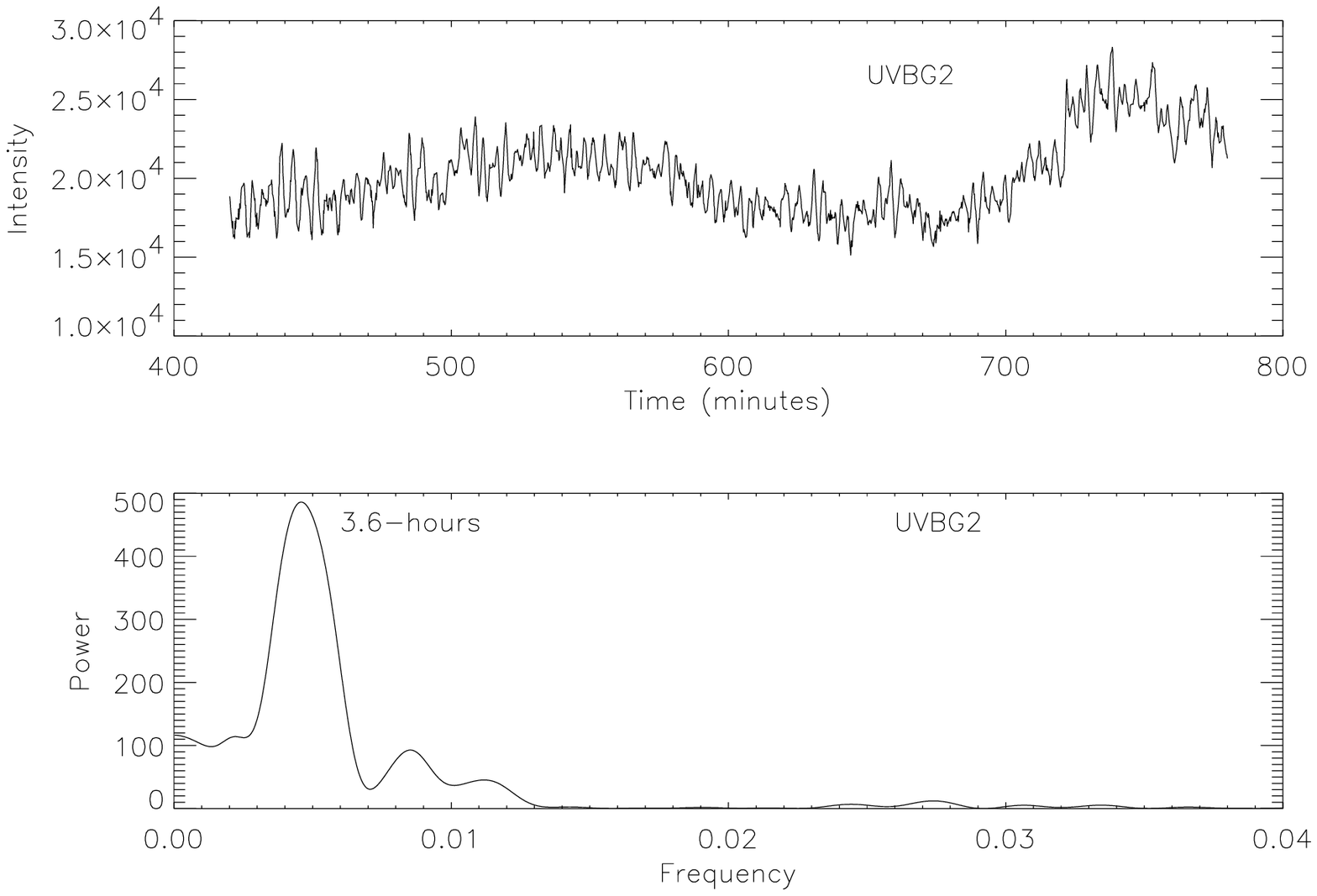}
\caption{{\it Left upper box:} An example of the light curve of a background region (UVBG1)
observed on May 24, 2003
(6 hours) with TRACE in 1600 $\AA$ UV continuum. {\it Left lower box:} The power spectra taken for
the light curve of the background region (UVBG1).
{\it Right upper box:}  The light curve of an another background region (UVBG2).
{\it Right lower box:} The power spectra taken for
the light curve of the background region (UVBG2).}
\label{FigVibStab}
    \end{figure}

We can summarize the main results derived from the analysis of 1600 $\AA$ continuum
observations are as follows:  (i)  The uv bright points, uv network elements and uv background
regions will exhibit a fluctuations with a smaller period in their intensity oscillations.
(ii)  We find evidence from the power spectrum analysis for a longer
period of oscillations: the uv bright points are associated with around 5.5 hours, the uv network
elements exhibit around 4.6 hours and whereas the background regions show around 3.4 hours.
(iii)  It is noted that the different features will have different period of intensity
oscillations and the reason for the existence of different periods is still to be investigated.
(iv) But, we can argue that the longer period of oscillations associated with all these three
features may be related to g-mode oscillations of the lower chromosphere.
These results confirm
the earlier findings that there is a signature of gravity waves in the chromosphere and
transition region derived from the analysis of time sequence of filtergrams and spectra
obtained in CaII H \& K, Mg b2 lines and from TRACE observations
(Dam\'e et al. 1984 and Kneer and von Uexkull, 1993, Rutten, and Krijger, 2003, Kariyappa,
et al. 2006).  We expect that the atmospheric gravity waves can exist in the stably stratified
photosphere and chromosphere, where convective overshoot is a natural mechanism to excite them.
For these atmospheric gravity waves no model has been proposed yet that is supported by all
the available observational constraints.


\begin{acknowledgements}

The time sequence of images and spectra have been obtained
simultaneously under Joint Observation Program (JOP163) with TRACE, SOHO/MDI and SOHO/CDS
experiments.  A part of the TRACE data analysis has been done
while one of us (Kariyappa) had CNRS Visiting Fellowship, France in 2003.  He will be
grateful to CNRS for the support and will be thankful to the Director and staff members of
Service d'A\'eronomie du CNRS for the kind hospitality \& support during his stay in Paris.

\end{acknowledgements}

\end{document}